# Microscopic Observation of Entangled Multi-Magnetoelectric Coupling Phenomenon

Sae Hwan Chun,[1] Kwang Woo Shin,[2] Kee Hoon Kim,[2] John F. Mitchell,[1] Philip J. Ryan,[3,4] and Jong-Woo Kim[3]

[1]Materials Science Division, Argonne National Laboratory, Argonne, IL 60439, U.S.A.
[2]CeNSCMR, Department of Physics and Astronomy, Seoul National University, Seoul 08826, South Korea
[3]Advanced Photon Source, Argonne National Laboratory, Argonne, IL 60439, U.S.A.
[4]School of Physical Sciences, Dublin City University, Dublin 11, Ireland

**Searching for new functionality in next generation electronic devices is a principal driver of material physics research. Multiferroics simultaneously exhibit electric and magnetic order parameters that may be coupled through magnetoelectric (ME) effects[1-11]. In single-phase materials the ME effect arises from one of three known mechanisms: inverse Dzyaloshinskii-Moriya (IDM) interaction[12-15], spin dependent ligand-metal (p-d) orbital hybridization[16-18], and exchange striction[19-21]. However, the coupling among these mechanisms remains largely unexplored despite envisioned potential capabilities. Here, we present cooperative tuning between both IDM interaction and p-d hybridization that leads to discrete ME states in $Ba_{0.5}Sr_{2.5}Co_2Fe_{24}O_{41}$. *In-situ* x-ray diffraction exposes the microscopic interplay between these two mechanisms, marked by a unique ME susceptibility upon electric and magnetic fields. The entangled multi-ME coupling phenomenon observed in this room-temperature ME hexaferrite offers a pathway to novel functional control for ME device applications.**

Many potential device applications such as multi-bit logic-memory[1-3], and spintronic devices with integration of electric polarizability[11] are based upon exploiting the functionality of ME effects. Intensive effort to achieve such functionalities has been made by searching for room-temperature multiferroics with large electric and magnetic polarizations. The prototypical system has been BiFeO$_3$ with ferroelectric polarization below $T_{FE}$ = 1100 K and antiferromagnetic order below $T_N$ = 643 K, which marks the highest ferroelectric polarization 9×10$^4$ μC/m$^2$ amongst the multiferroics[1,7,22]. Due to the IDM interaction, the magnetic easy plane is determined by the electric polarization (*P*) and the antiferromagnetic ordering vector[23,24], and thereby the magnetic domain structure can be controlled by an electric field (*E*) (Ref. 23). The magnetic field (*H*) control of *P*, however, is only available at relatively high field (~160 kOe) because of small magnetic susceptibility of the antiferromagnetic behavior[22]. Various efforts continue to enhance the ME coupling strength (d*P*/d*H* ~55 ps/m), for example, by thin film engineering[1] or chemical doping[25].

The remarkably large ME susceptibility of ~3200 ps/m at room temperature in hexaferrites was also regarded as a breakthrough towards the realization of a potential ME device[26-29]. In particular, the (Ba,Sr)$_3$Co$_2$Fe$_{24}$O$_{41}$ (Co$_2$Z-type) hexaferrite exhibits control of *P* by *H* (Ref. 28) and *M* by *E* (Ref. 29) under low *H* ~0.1 kOe at room temperature. These ME effects are observed in a transverse conical magnetic state (Fig. 1b) that hosts the IDM interaction[27,28]. This ME mechanism yields an antisymmetric *P* response upon switching *H* as the magnetic chirality $\vec{\mu}_L \times \vec{\mu}_S$ is reversed in the relation $\vec{P}_{asym} \propto \sum \vec{k}_0 \times (\vec{\mu}_L \times \vec{\mu}_S)$, where $\vec{k}_0$ is the magnetic modulation vector, and $\vec{\mu}_L$ and $\vec{\mu}_S$ are the net moments of the magnetic *L* and *S* blocks, respectively[12-15]. However, a characteristic ME feature of the Co$_2$Z-type hexaferrite is a symmetric *P* behavior while showing a small contribution from $P_{asym}$ (Ref. 29), which indicates the importance of another ME mechanism[28,29]. When more than one ME mechanism is present, an interplay between individual ME elements is expected. For instance, the IDM interaction and the p-d hybridization can influence each other due to their connection through the spin-orbit coupling. As such couplings interact, a new functional control of the ME state by tuning this interplay is possible. In this work, we look into this potential through observing the structural response to the ME effect under *E* and *H* by *in-situ* x-ray diffraction on the Co$_2$Z-type hexaferrite.

The symmetric ME behavior of the Co$_2$Z-type hexaferrite results from either p-d hybridization or exchange striction due to the quadratic nature of these mechanisms. Assessing the crystal structure (Fig. 1a), the p-d hybridization is considered the primary source of the symmetric ME behavior. In particular, we note the unit comprised of the corner-shared octahedron and tetrahedron at the boundary of *L* and *S* blocks, which contains off-centered Fe/Co ions (depicted inside the dashed lines in Fig. 1a). This off-centering can give rise to asymmetric hybridization of the d-orbitals in the transition metal (TM) ions with the p-orbitals in the surrounding ligand oxygen ions. Subsequently a local electric polarization forms $\vec{P}_{pd} \propto \sum_i (\vec{e}_i \cdot \vec{S})^2 \vec{e}_i$, where $\vec{S}$ and $\vec{e}_i$ are the spin and the vector connecting the TM-O bond, respectively[16-18]. The asymmetric hybridization due to the off-centering leads the direction of $P_{pd}$ to be determined by the relationship between the TM ion and its nearest oxygen triangular layer. For instance, when the spin is parallel to one edge of this oxygen triangle, the in-plane component of $P_{pd}$ points towards this edge (top panel of Fig. 1d). On the other hand, when the spin is perpendicular to a particular edge, the in-plane component points towards the apex opposite to the edge (bottom panel of Fig. 1d). The out-of-plane component of $P_{pd}$ is along the off-centered direction.

Considering the crystallographic symmetry and the spin structure of the transverse conical state in the Co$_2$Z-type hexaferrite, we find that the sum of all the local $P_{pd}$ over the unit cell can result in a finite, net in-plane polarization $P_{sym}$. This $P_{sym}$ under $H$ is effectively described by the sum of $P_{pd}$ in the octahedra at the boundaries between *L* and *S* blocks, whose TM ions are most off-centered along the *c*-axis. Figures 1e&f represent these octahedra along with their spin and electric polarization ($P_{pd}$). For the in-plane $H$ along $[\bar{1}10]$ (or identical directions of three-fold rotational symmetry $[1\,0\,0]$ and $[0\,\bar{1}\,0]$), the sum of $P_{pd}$ (|| $P_{sym}$) is perpendicular to the $H$ direction (Fig. 1e). When $H$ is rotated by 90°, the sum of $P_{pd}$ becomes reversed (Fig. 1f). The magnetic field dependence of the electric polarization shows sign reversal of $P$ (Fig. 1c), verifying p-d hybridization (Fig. 1d). Exchange striction is also able to produce symmetric $P$ vs. $H$ behavior, but in this case, $P$ would not change with the rotation of $H$ since the spin alignment of both parallel and antiparallel components would be preserved.

Both distinct ME mechanisms, i.e. the IDM interaction and the p-d hybridization arising at the boundary of *L* and *S* blocks, can couple to each other. The $P$ driven by the IDM interaction emerges as the antisymmetric spin exchange in the non-collinear spins shifts the ligand oxygen ion[15,30]. The $P$ vs. $H$ measurement detects the macroscopic $P$ ($P_{sym}$ and $P_{asym}$), but not necessarily the nature of their correlation. In contrast, the x-ray diffraction intensity varies upon local structural distortions resulting from the interplay between the two ME mechanisms. Therefore, *in-situ* x-ray diffraction under $E$ and $H$ is suitable in order to reveal any such intrinsic microscopic coupling behavior.

The P6$_3$/mmc symmetry is broken due to small atomic displacements, verified by observing clear Bragg peak intensities at the (0, 0, $l$) with $l$ = odd integers, which are forbidden for this space group. We find experimentally that the intensities of the forbidden (0, 0, 9) and (0, 0, 19) reflections depend on the ME effect, demonstrating the structural response to $H$. These two Bragg peaks exhibit changes of over 10 % (Fig. 2 and Supplementary Fig. S1) whereas other (0, 0, $l$) peaks display smaller changes less than 2 % (see Supplementary Fig. S2). The intensity of the (0, 0, $l$) reflections depends on ME response of the oxygen layer distortions at the boundary of $L$ and $S$ blocks. Greater intensity variation arises in certain $l$ values corresponding to the inter-layer distances. For instance, the oxygen layers located at c/9, c/16, and c/36 (inside the dashed lines in Fig. 1a) and their symmetric counterparts are separated by c/4.5, c/8, and c/18, respectively. Therefore, the (0, 0, 9) and (0, 0, 19) peaks are most susceptible to the ME effect.

The intensity of (0, 0, 9) reflection for a single ME state is monitored under different $H$ conditions. This state (referred as (+,+) state) is prepared by a ME poling procedure that applies transverse poling fields $+E_P$ and $+H_P$ along [100] and [$\bar{1}\bar{2}0$] directions, respectively (Fig. 2a and see Methods). For $H$ applied along the initial poling direction ($\Psi = 0°$), the intensity decreases symmetrically with $H$, as shown in Fig. 2d. As the azimuthal $H$ direction is rotated, it displays a prominent antisymmetric field dependence at $\Psi = 30°$ followed by symmetric dependence again at $\Psi = 60°$ (Figs. 2e&f). When $H$ is counter-rotated from the initial direction the intensity shows reversed antisymmetric behavior at $\Psi = -15°$. This azimuthal $H$ dependence is also observed in the (0, 0, 19) reflection (see Supplementary Fig. S1), while the intensity variation is reversed with respect to the (0, 0, 9) reflection due to the out of phase contribution to the structure factor from the combination of ionic displacements.

This azimuthal $H$ dependence is understood by the rearrangement of $P_{sym}$ and $P_{asym}$ with the rotation of $H$. With varying $\Psi$, the interplay between these two $P$ components produces different intensity dependencies. The left panel of Fig. 2h depicts the initial ME state at $\Psi = 0°$, where $P_{sym}$ has a dominant component parallel to $+H_P$ while $P_{asym}$ points along $+E_P$. In this configuration, $H$ as well as $P_{sym}$ in the *ab*-plane is along one of the three-fold symmetric directions of the P6$_3$/mmc space group, and $P_{asym}$ is perpendicular to this direction. Consequently, $P_{asym}$ lies along the direction normal to the crystallographic mirror symmetry plane spanned by $P_{sym}$ and the *c*-axis, so that the atomic distortion should be equivalent regardless of the $P_{asym}$ reversal. Since the IDM interaction maintains $P_{asym}$ perpendicular to the net magnetic moment, the azimuthal $H$ rotation leads $P_{asym}$ to rotate with the net magnetic moment. Meanwhile, $P_{sym}$ arising from the p-d hybridization rotates twice as much and opposite to the $H$ rotation due to the relationship between the spin and the TM-O direction (for details, see Supplementary Fig. S3). At $\Psi = 30°$, $P_{sym}$ and $P_{asym}$ become parallel and antiparallel upon switching $H$ (Fig. 2i). In this particular arrangement, the coupling between $P_{sym}$ and $P_{asym}$ yields an asymmetric ionic displacement leading to the antisymmetric $H$ dependence. Similarly, the intensity variation at $\Psi = -15°$ exhibits reversed antisymmetric behavior (Fig. 2g). The symmetric intensity variation is observed at every 60° which serves as fiducial angles where the sign of the antisymmetric behavior switches. The rearrangement of $P_{sym}$ & $P_{asym}$ by the $H$ rotation repeats with a period of 120° due to the hexagonal symmetry (Fig. 2f).

Reversed $H$ intensity dependence for the different initial ME states further establishes that our observations arise from a microscopic structural response. For the ME state, (-,+), which is prepared by reversing the direction of poling $E$ (Fig. 3c), both (0, 0, 9) and (0, 0, 19) reflections exhibit the reversal of the antisymmetric behaviors observed in the (+,+) state at $\Psi = 30°$, as shown in Figs. 3a&b. This result confirms that the initial ME poling determines the microscopic correlation between the multi-ME couplings (Fig. 3c). The different arrangement of $P_{sym}$ and $P_{asym}$ for each ME state leads to the distinct, reversed responses of the symmetry modification.

The initial ME poling establishes the magnetic chirality of the transverse conical magnetic order. While the directions of $\vec{\mu}_{L1}$ and $\vec{\mu}_{L2}$ within the *ab*-plane are coupled with $P_{sym}$, the formation of both *c*-axis components of $\vec{\mu}_{S1}$ and $\vec{\mu}_{S2}$ is associated with $P_{asym}$ (Fig. 1b). As a result, chirality is determined by the initial poling, and the magnetic state attains its own ME characteristics through the arrangement of $P_{sym}$ and $P_{asym}$, not achievable for each individual ME mechanism alone.

One example of such functional ME characteristics is found under simultaneous application of $E$ and $H$. Figure 4 shows the external $E$ response of the $H$-induced intensity variations of the (0, 0, 9) peak at $\Psi = 30°$. The intensity change between $E = +2$ MV/m and -2 MV/m is observed only at $-H$, while the intensity at $+H$ remains unchanged. This result is attributed to the entanglement of the magnetic chirality, $P_{sym}$, and $P_{asym}$. The right inset of Fig. 4 depicts the arrangement of $P_{sym}$ and $P_{asym}$ being parallel at $+H$. In order to tilt to the external $E$ field, both $P_{sym}$ and $P_{asym}$ should rotate toward the same direction. In fact, this is forbidden by the inter-relationship between the p-d hybridization and IDM interaction, coupled through the magnetic moment. In contrast, the antiparallel polarization configuration (left inset) at $-H$ leads the $P_{sym}$ and $P_{asym}$ to rotate in opposite directions tilting towards $E$ maintaining their coupling constraint. This rotational pathway results in a higher ME susceptibility to $E$ under $-H$, as indicated by the reflection intensity change.

Our observation of the entangled multi-ME coupling phenomenon in the $Co_2Z$-type hexaferrite demonstrates a novel example of a structural response driven by the interplay of the IDM interaction and the p-d hybridization. The relationship of the local $P$ and the corresponding structural distortions established in this work illuminates a route to control the linear and quadratic ME effects via their correlated coupling in the hexaferrite. The discrete ME susceptibility upon the arrangement of $P_{sym}$ and $P_{asym}$ offers a ME functionality that solely depends on the interplay of the two ME mechanisms. The arrangement persists even at room temperature (see Supplementary Fig. S4), which can impact novel functional control for the ME device applications.

**Single crystal growth**

The Co$_2$Z-type hexaferrite Ba$_{0.5}$Sr$_{2.5}$Co$_2$Fe$_{24}$O$_{41}$ single crystals were grown from the Na$_2$O-Fe$_2$O$_3$ flux in air[29]. The chemicals were mixed with the molar ratio of BaCO$_3$ : SrCO$_3$ : CoO : Fe$_2$O$_3$ : Na$_2$O = 1.97 : 17.72 : 19.69 : 53.61 : 7.01. The mixed chemicals were melted at 1420˚C in a platinum crucible and cooled down to room temperature after several thermal cycles between 1080˚C and 1420˚C. The grown crystals were heat-treated at 900˚C under flowing O$_2$ gas for 8 days to remove oxygen vacancies. The single crystals were pre-screened by comparing the magnetization to the one in Ref. 29. The Co$_2$Z-type phase was confirmed by the (0, 0, $l$) intensity profile that shows the Bragg peaks only associated with the phase (see Supplementary Fig. S5).

**Magnetoelectric current measurement**

The sample was cut into a rectangular parallelepiped shape with the largest surface (1.5 mm$^2$) normal to an arbitrary in-plane direction, and then the electrodes were prepared on the parallel surfaces with a silver epoxy. The sample was placed in Physical Property Measurement System (PPMS$^{TM}$, Quantum Design, USA) and its ME current was measured by using a current meter while sweeping $H$. Before the measurement, an $E$ field poling procedure was performed to prepare a single ferroelectric domain phase: $E$ = 100 kV/m was applied to the sample while sweeping an in-plane $H$ from 30 kOe to 2 kOe, i.e. from outside to inside the ferroelectric phase, in a configuration where $H$ is transverse to the poling $E$. The ME current was recorded in $H$ sweep from 2 kOe to -30 kOe and then integrated to determine the electric polarization. For another configuration, where the in-plane $H$ is rotated by 90˚, the sample was rotated since the $H$ direction of the PPMS is fixed. Before rotating the sample, the same $E$ poling procedure was performed to prepare the same initial state. The ME current for this rotated configuration was measured in the same manner.

**Preparation of single magnetoelectric domain states**

The ME poling was carried out to realize a single ME domain states with distinct electric and magnetic polarization directions. The (+,+) state was prepared by poling $E$ and $H$ towards [100] and [$\bar{1}\bar{2}0$], respectively, as shown in Fig. 2a. Application of the poling $E$ (= 300 kV/m) was started at $H$ = 30 kOe outside the ferroelectric phase in the $H$ region < 22 kOe at 30 K and kept while the poling $H$ was reduced to zero inside the ferroelectric phase. For the different (-,-) state, the poling $E$ was reversed towards $[\bar{1}00]$ in the ME poling process.

### *In-situ* single crystal x-ray diffraction

X-ray single crystal diffraction was performed at the 6-ID-B beamline of the Advanced Photon Source. The sample with the electrodes was mounted on the cold finger of a closed cycle helium refrigerator. For the *in-situ* x-ray diffraction study under different $H$ and $E$, an electromagnet was placed beside the cold finger and the refrigerator was modified to allow the wire connection between the electrodes and the external voltage source. Incident x-ray energy in the range of 7.06 - 7.13 keV was selected for each Bragg peak after checking the energy dependence of the intensity to avoid multiple-beam diffraction (see Supplementary Fig. S6). The intrinsic structural change under $E$ or $H$ fields was monitored by the integrated intensity from the rocking curve at each Bragg peak.

**Acknowledgements**

Work in the Materials Science Division of Argonne National Laboratory (sample characterization and data analysis) was supported by the U.S. Department of Energy (DOE), Office of Science, Basic Energy Sciences, Materials Science and Engineering Division. Use of the Advanced Photon Source, an Office of Science User Facility operated for the U.S. DOE Office of Science by Argonne National Laboratory, was supported by the U.S. DOE under Contract No. DE-AC02-06CH11357. Work in Seoul National University (crystal growth) was supported by Creative Research Initiatives (2010-0018300).

**Author contribution**

S.H.C. and J.-W.K. conceived the project. S.H.C., P.J.R., and J.-W.K. performed the experiment. K.W.S grew the single crystals under supervision of K.H.K. S.H.C. and J.-W.K. analyzed the data. All authors discussed the results. S.H.C., P.J.R. and J.-W.K. led the manuscript preparation with contributions from all authors.

**Author information**

The authors declare no competing financial interests. Correspondence should be addressed to S.H.C. (pokchun81@gmail.com) or J.-W.K. (jwkim@aps.anl.gov).

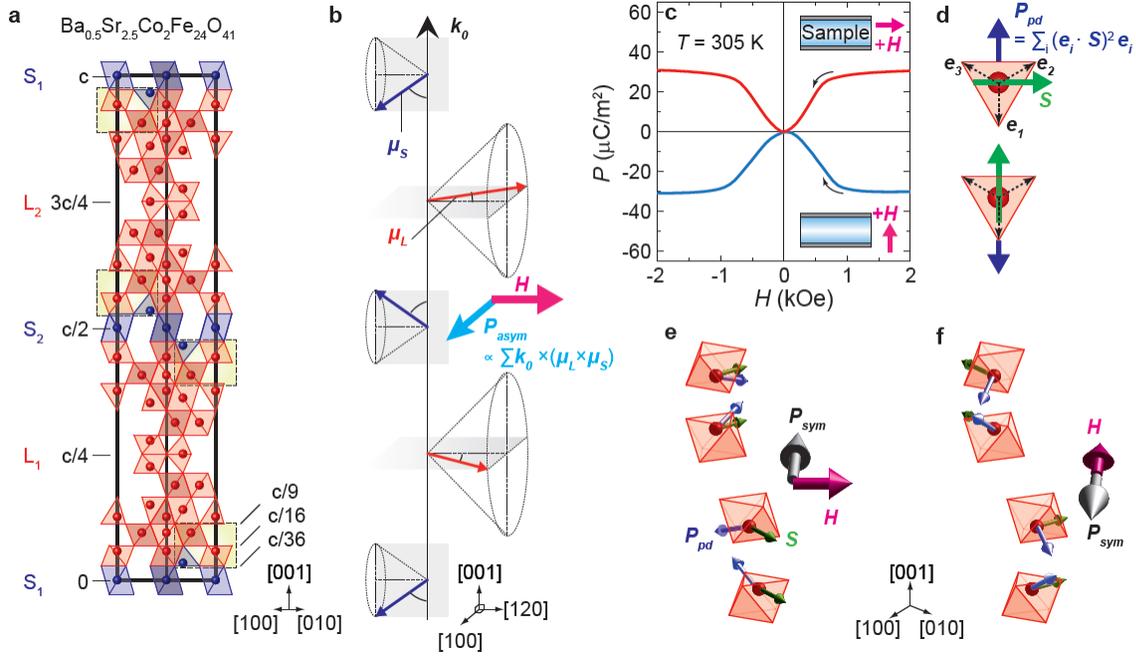

**Figure 1 | Two origins of direct ME effect realized in the Co₂Z-type hexaferrite.** (a) Crystal structure with the representation of magnetic *S* (blue) and *L* (red) blocks. Dashed box denotes two octahedra and a tetrahedron consisting of Fe/Co ions and ligand oxygen ions at the boundary of *S* and *L* blocks, whose competing superexchanges yield non-collinear spin structures. (b) Transverse conical magnetic order represented by the net magnetic moments of *S* and *L* blocks ($\vec{\mu}_S$ and $\vec{\mu}_L$) with the magnetic modulation vector $\vec{k}_0$ = (0, 0, 1). The dotted cones are guide lines. $P_{asym}$ is sum of the electric polarization induced by IDM interaction and has the in-plane net component perpendicular to the conical axis defined by *H* direction. (c) In-plane electric polarization vs. *H* curve (red) for the transverse in-plane *H* configuration (top inset) and the curve (blue) for the longitudinal in-plane *H* configuration (bottom inset) for a ME poled state. The ME poling was first performed under the transverse configuration by applying +*E* poling, followed by the ME current measurement in the same or the 90° rotated configurations. (d) $P_{pd}$ arising from the relationship between the spin and the vectors connecting the TM-O bonds. (e,f) The collection of the octahedron with the most off-centered Fe/Co (between z = c/36 and c/16) and their symmetry-equivalent at the boundary of *S* and *L* blocks. Green, and blue arrows denote the spin and $P_{pd}$ in the individual octahedron, respectively. Sum of $P_{pd}$ results in in-plane $P_{sym}$ that can be reversed by 90° rotation of *H* field.

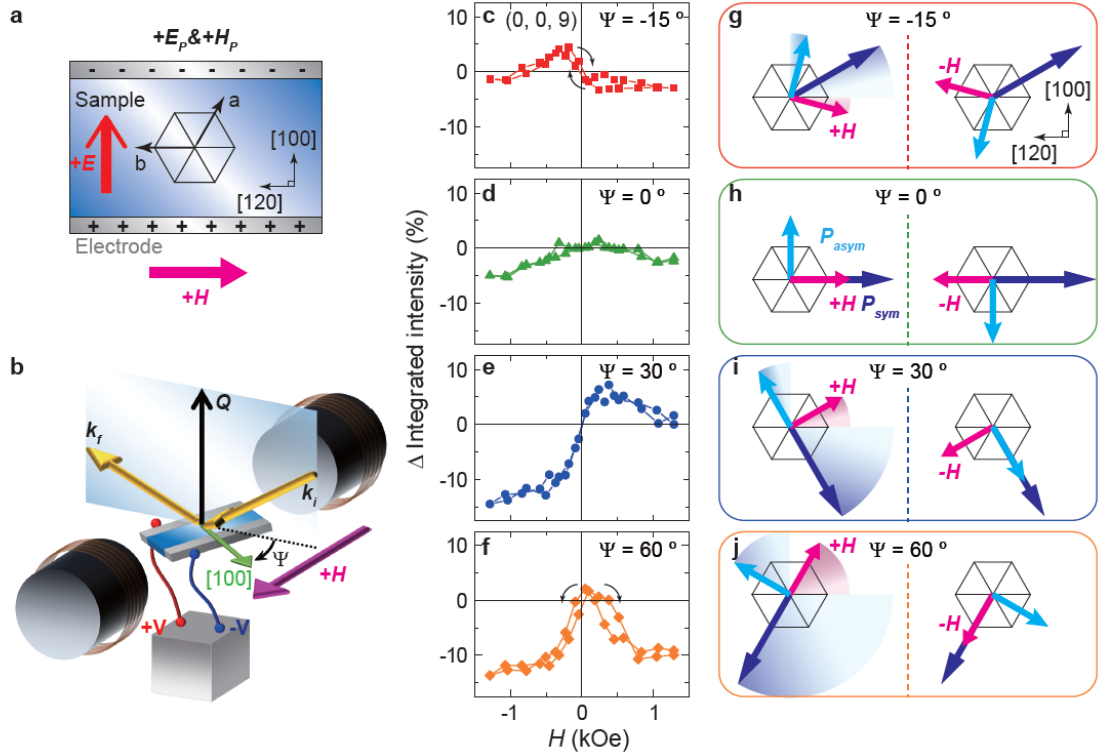

**Figure 2 | Azimuth angle dependence of the (0, 0, 9) intensity variations upon switching magnetic field.** (a) Illustration of applying $E$ and $H$ fields to prepare a single ME state, (+,+). (b) The x-ray diffraction geometry implementing the vertical scattering plane defined by the incident ($k_i$) and out-going ($k_f$) wave vectors with the $H$ application normal to the plane. (c-f) Integrated intensity variations of the (0, 0, 9) peak upon switching $H$ field for azimuthal angles $\Psi = -15°$, $0°$, $30°$, and $60°$. $\Psi$ is defined by the angle between [100] and the scattering plane. (g-j) The possible rearrangement of $P_{sym}$ and $P_{asym}$ at $+H$ (left panel) and $-H$ (right panel) for each $\Psi$. The relationship of $\vec{P}_{asym} \propto \sum \vec{k}_0 \times (\vec{\mu}_L \times \vec{\mu}_S)$ leads $P_{asym}$ to concurrently rotate with $H$ while maintaining a perpendicular relationship and the handedness with the magnetic chirality. On the other hand, the relationship of $\vec{P}_{pd} \propto \sum_i (\vec{e}_i \cdot \vec{S})^2 \vec{e}_i$ forces $P_{sym}$ to rotate opposite to and by twice as much as the $H$ rotation.

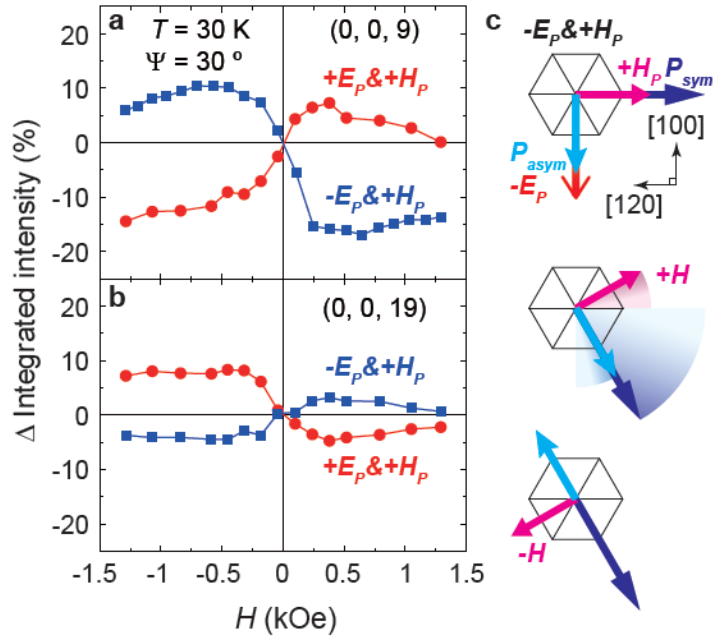

**Figure 3 | Reversal of *H* dependent structural distortion upon -*E* poling.** (a, b) The (0, 0, 9) and (0, 0, 19) intensities presented at $\Psi = 30°$ and their antisymmetric behaviors between (+,+) and (+,-) ME states. (c) The initial configuration of $P_{sym}$ and $P_{asym}$ for the (-,+) state prepared at $\Psi = 0°$ (top). The parallel and antiparallel rearrangements of the polarizations realized at $\Psi = 30°$ for +*H* (middle) and -*H* (bottom), respectively. These arrangements are opposite to the (+,+) state as depicted in Fig. 2i.

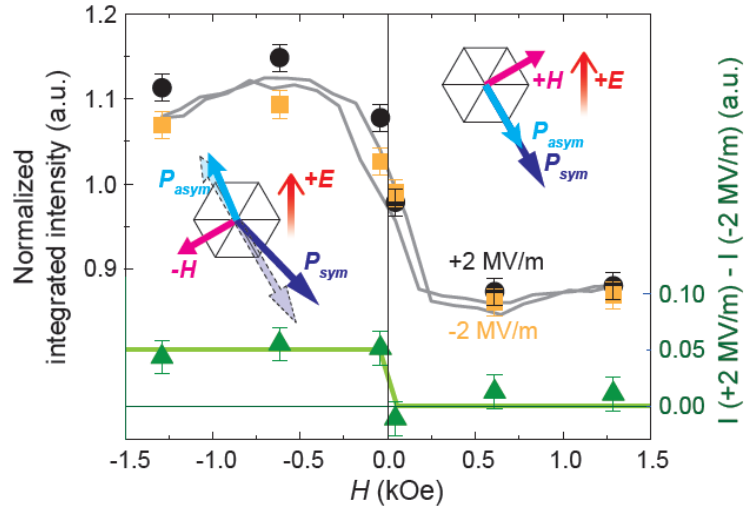

**Figure 4 | Structural response by simultaneous application of *E* and *H*.** The measurement was carried out for (0, 0, 9) peak with the (-,+) state at $\Psi = 30°$ and $T = 30$ K. The solid black circle and orange square denote the integrated intensities measured under application of $E = +$ 2MV/m and $-$ 2MV/m, respectively (left scale). Their difference is represented by the solid green triangle (right scale) exhibiting the notable change under -*H*. The solid grey curve represents the *H* dependence at $E = 0$ MV/m. All data is normalized by the intensity at $H = 0$ kOe and $E = 0$ MV/m. The left and right insets illustrate the antiparallel and parallel arrangements of $P_{sym}$ and $P_{asym}$ at -*H* and +*H*, respectively, under +*E*. The dashed arrows in the left inset depict the initial position of the polarizations at $E = 0$ MV/m. Application of +*E* directs $P_{sym}$ and $P_{asym}$ to rotate opposite to each other, which is a favorable pathway to follow the *E* field.

# Supplementary Information

## A.  Azimuth angle dependence of the (0, 0, 19) intensity variations

The intensity of the (0, 0, 19) reflection anticorrelates with that of the (0, 0, 9). The symmetric and antisymmetric structural responses upon the azimuthal $H$ direction is also notable in the (0, 0, 19) intensity variations. Figure S1 summarizes the intensity variations at different azimuth angles $\Psi$. The intensity variation at $\Psi = -15°$ shows an antisymmetric behavior, while at $\Psi = 0°$ much reduced antisymmetric or enhanced symmetric behaviors are found. The intensity variation at $\Psi = 30°$ exhibits the reversed antisymmetric behavior, which becomes symmetric at $\Psi = 60°$. This azimuthal $H$ direction dependence is consistent with the one observed in the (0, 0, 9).

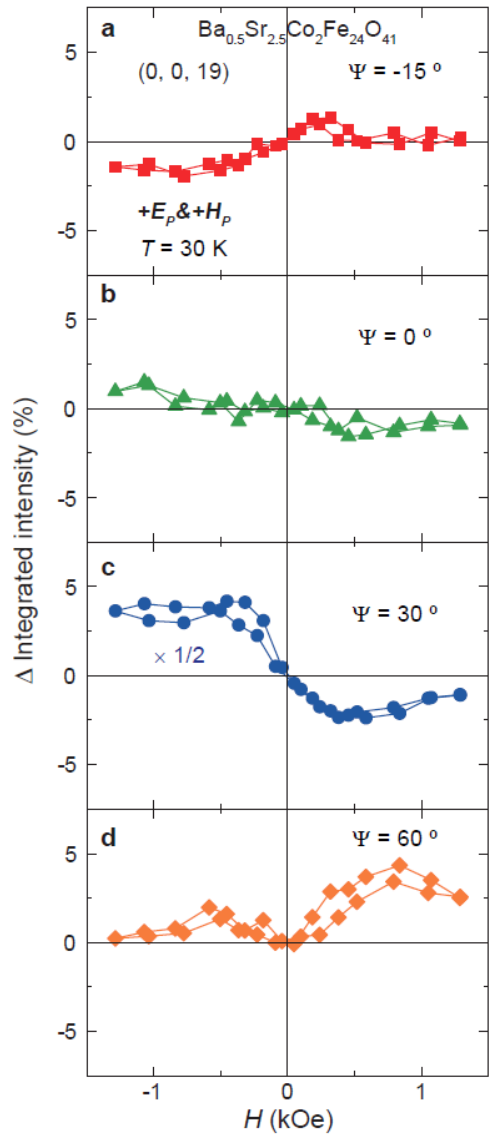

**Figure S1 | Azimuthal angle dependence of the (0, 0, 19) intensity variations upon switching magnetic field.** (a-d) Variations of the integrated (0, 0, 19) intensity upon switching $H$ for azimuthal angle $\Psi$ = -15°, 0°, 30°, and 60°. The data at the $\Psi$ = 30° is reduced by half for comparison with those at other $\Psi$ angles.

### B. Intensity variations of other Bragg peaks

The Bragg peaks we accessed other than the (0, 0, 9) and (0, 0, 19) peaks also show intensity variation under switching the $H$ field. In contrast to the (0, 0, 9) and (0, 0, 19), which display a substantial total intensity change exceeding 10 %, the change in other Bragg peaks is less than 2 % in the configuration where the most pronounced change is observed (Fig. S2).

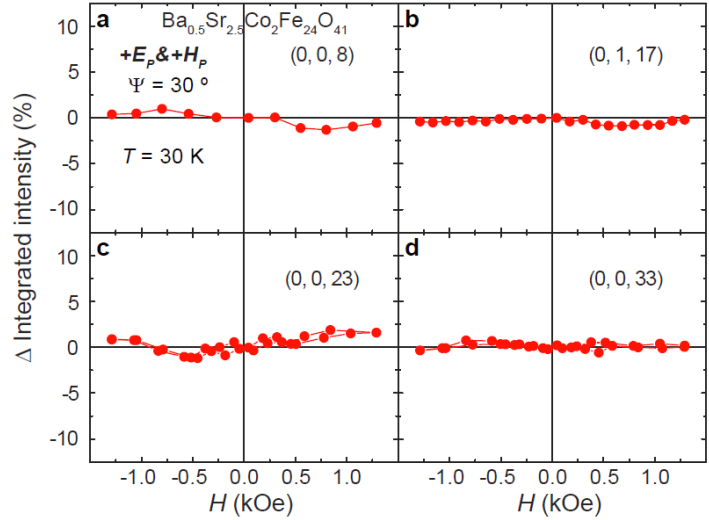

**Figure S2 | Magnetic field dependence of the Bragg peak intensities.** (a-d) The (0, 0, 8), (0, 1, 17), (0, 0, 23), and (0, 0, 33) peak intensities at $T = 30$ K with the azimuthal $H$ direction set at $\Psi = 30°$ for the (+,+) state.

## C. Rotational relation between the spin and the electric polarization driven by the p-d hybridization

The spin rotation at the transition metal (TM) ion with respect to the surrounding oxygen octahedron leads to a distinct degree of the p-d hybridization along the TM-O bond directions. For the off-centered octahedron, the p-d hybridization with three nearest oxygen ions dominates over those of the other three more distant ions, such that the resultant electric polarization ($P_{pd}$) is effectively described by the relationship between the TM ion and the nearest oxygen triangular layer. A simple way to understand the in-plane component of $P_{pd}$ upon the spin rotation is to examine the spin ($S$) and the TM-O bond ($e_1$, $e_2$, and $e_3$) vectors projected on the oxygen layer, as shown in Fig. S3. When $S$ is parallel to the top edge of the triangle (the far left panel), $P_{pd} \propto \sum_i (e_i \cdot S)^2 e_i$ has no contribution from the $e_1$ component as $e_1 \cdot S = 0$. Thus, the vector sum of the $e_2$ and $e_3$ components yields $P_{pd}$ bisecting $e_2$ and $e_3$. In the case that $S$ is rotated by 30° counter-clockwise (the second panel), $S$ becomes parallel to $e_2$ and thereby $P_{pd}$ is dominated by the $e_2$ component, $P_{pd}$ having rotated by 60° clockwise. The in-plane $P_{pd}$ direction upon further rotation of $S$ is determined in the same manner, which reveals that $P_{pd}$ counter-rotates to $S$ by twice that of the $S$ rotation.

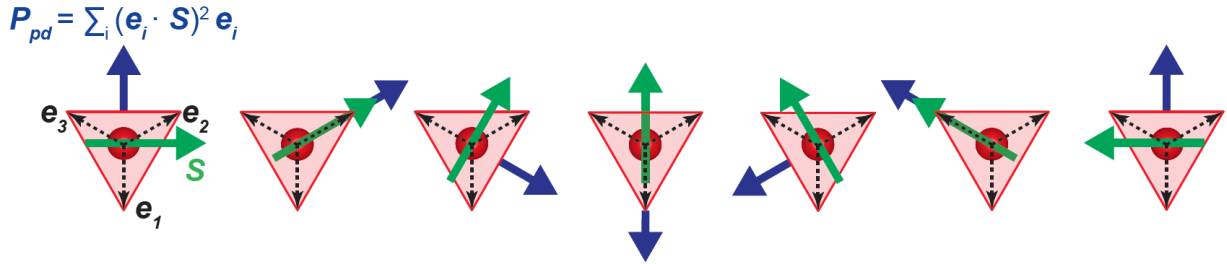

**Figure S3 | Change of the in-plane $P_{pd}$ component upon the spin rotation.** The red triangle depicts the face of the oxygen octahedron that is closest to the Fe/Co ion (red sphere). The dashed black arrows denote the $e_1$, $e_2$, and $e_3$ vectors that connect the Fe/Co ion to the oxygen ions on the face. The blue and green arrows represent the in-plane $P_{pd}$ component and the spin, $S$, respectively.

## D. Temperature dependence of the (0, 0, 9) intensity variation

The antisymmetric variation of the (0, 0, 9) peak intensity upon switching the magnetic field ($H$) is observed at higher temperature, including 300 K (Fig. S4). The amount of intensity variation at 300 K is smaller than those at 173 K as well as 30 K, indicating that the magnetically induced structural distortion is reduced. This result is consistent with ferroelectricity existing below ~410 K and the electric polarization decreasing as temperature approaches the transition temperature.

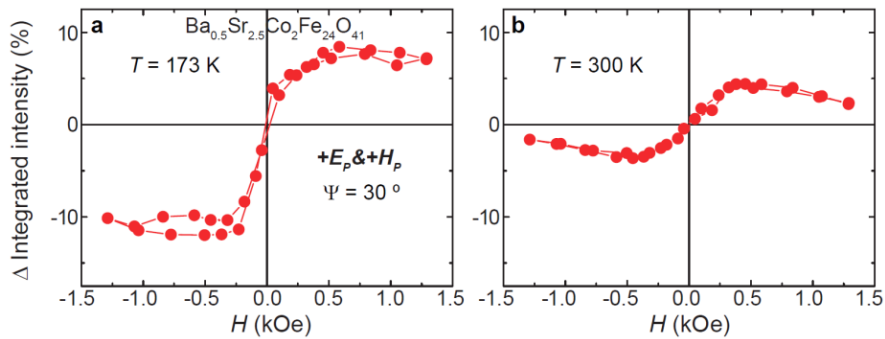

**Figure S4 | The (0, 0, 9) peak intensity variation at higher temperature.** The antisymmetric change of the integrated intensities at $\Psi = 30°$ for the (+,+) state at (a) $T = 173$ K and (b) 300 K.

### E. Sample characterization

A specular diffraction (0, 0, $l$) scan in Fig. S5 displays the Bragg peaks at the $l$ = integers which are associated only with the Co$_2$Z-type hexaferrite (a = 5.8614 Å and c = 51.8878 Å at $T$ = 30 K). The peak intensities at the (0, 0, odd) have not been reported prior to this work. The intense incident x-ray beam in the synchrotron facility allows us to resolve these intensities which are smaller by factors of $10^3$-$10^5$ than those at the (0, 0, even). The appearance of these (0,0,odd) reflections, forbidden in space group P6$_3$/mmc, implies that the crystallographic symmetry is lowered. Maximal subgroups compatible with observed reflection conditions include $P\bar{3}m1$ or $P\bar{6}m2$, the latter being noncentrosymmetric. Further structural work will be required to establish unequivocally the structure of the Co$_2$Z-type hexaferrite.

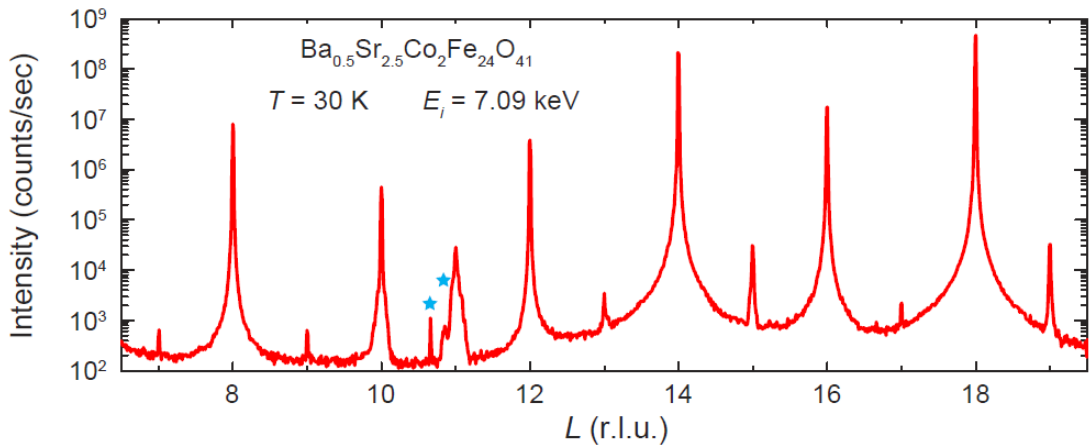

**Figure S5 | The (0, 0, $l$) profile of the Co$_2$Z-type single crystal.** The data were collected at $T$ = 30 K by employing the incident x-ray energy $E_i$ = 7.09 keV. The blue stars mark the peaks coming from other sources in the instrument.

### F. Incident x-ray energy dependence of the (0, 0, 9) peak intensity

The (0, 0, odd) peaks exist regardless of the incident x-ray energy ($E_i$). Figure S6 shows the $E_i$ dependence of the (0, 0, 9) peak as an example. The enhanced intensity at certain $E_i$ values is attributed to multiple reflection. Under the application of $H$, the intensity profile remains nearly unchanged except for the constant offsets in the monitored $E_i$ range. This ensures that the $H$ induced intensity variation arises from the intrinsic structural distortion, not from a change in multiple reflection. The $E_i$ values separated from the multiple reflection were chosen for investigating the intensity variations of the (0, 0, odd) peaks upon the $H$ and $E$ fields.

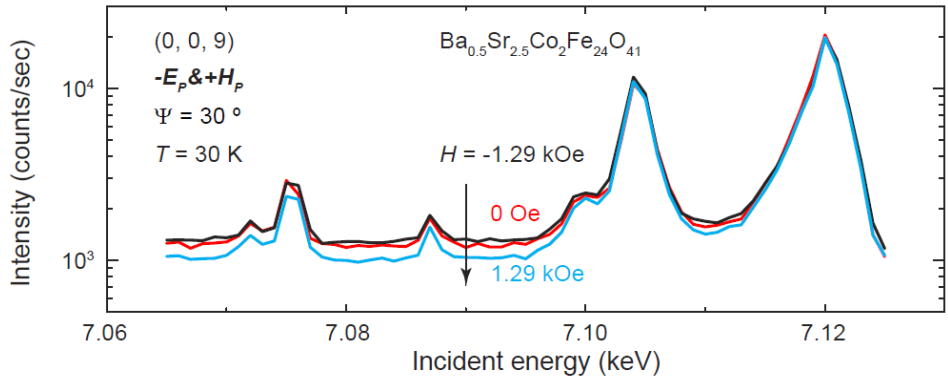

**Figure S6 | Incident x-ray energy dependence of the (0, 0, 9) peak intensity.** The black, red, and blue curves represent the data at $T = 30$ K under the application of $H = $ -1.29, 0, 1.29 kOe, respectively, at $\Psi = 30°$ for the (-,+) state.